\documentclass[preprint,12pt]{elsarticle}

\usepackage{amssymb}
\usepackage{color}
\usepackage{graphicx}

\journal{The European Physical Journal C}

\begin{document}

\begin{frontmatter}



\title{Is the mysterious {\it Punctum} a primordial black hole?}


\author{Man Ho Chan}

\address{Department of Science and Environmental Studies, The Education University of Hong Kong, Tai Po, New Territories, Hong Kong, China}

\ead{chanmh@eduhk.hk}

\begin{abstract}
A recent study has discovered a highly polarized millimeter continuum source called {\it Punctum} located at the central region of NGC 4945 which doesn't fit any known category. Apart from the radio detection at $\approx 100$ GHz, there is no identified counterpart to {\it Punctum} in other radio, infra-red, optical, and X-ray bands. All available potential models, including magnetar, supernova remnant and microquasar, cannot provide satisfactory explanation for the exceptionally high polarization of $\sim 50$\% and large luminosity in mm band $L_{\rm mm} \approx 2\times 10^{35}$ erg/s within a small size $<2$ pc shown in {\it Punctum}. In this article, we discuss the possibility that {\it Punctum} is indeed a primordial black hole (PBH) with a dark matter density spike. We show that the synchrotron radiation due to the electrons and positrons produced from dark matter annihilation can satisfactorily explain the observed properties of {\it Punctum} and we can effectively obtain the annihilation cross section $\langle \sigma v \rangle \sim 10^{-33}$ cm$^3$/s using our developed analytic framework. Furthermore, if the {\it Punctum} PBH mass is $\sim 10-100M_{\odot}$, the preferred dark matter mass range is $m_{\rm DM} \sim 10-1000$ MeV. 
\end{abstract}


\end{frontmatter}



\section{Introduction}
Recent millimeter-wave observations in the $\sim 30-300$ GHz range have revealed many results which have renewed our understanding about magnetic fields of galaxies and gas dynamics \cite{Decarli,Morii,Guerra,Roo}. In particular, one recent observation from the Atacama Large Millimeter/submillimeter Array (ALMA) has discovered a mysterious source which emits significant radio signals in the mm band \cite{Shablovinskaia}. Over the 14-day interval between two observations, the total radio fluxes detected are $0.104\pm 0.018$ mJy and $0.125\pm 0.016$ mJy respectively at frequencies $\nu=90.5-104.5$ GHz, which correspond to a large radio luminosity $L_{\rm mm} \approx 2 \times 10^{35}$ erg/s \cite{Shablovinskaia}. Moreover, the source exhibits an exceptionally high level of polarization of 51\%. Since the source was unresolved, the size should be $<2$ pc ($<0.11$ arcsec), which refers to as {\it Punctum} due to its compactness and unknown origin \cite{Shablovinskaia}. Such a high level of polarization suggests that the radio signals likely originate from synchrotron radiation with a highly uniform magnetic field structure. However, it is mysterious that there is no counterpart to {\it Punctum} in other radio, infra-red, optical, and X-ray bands. 

Several possible models have been highlighted to account for the emission of {\it Punctum}. For example, magnetars can produce radio emission with high polarization \cite{Liu}. However, the radio luminosity of two known magnetars SGR J1745-2900 and XTE J1810-197 only produce $\sim 5\times 10^{31}$ erg/s \cite{Torne} and $\sim 2\times 10^{30}$ erg/s \cite{Camilo} respectively, which are at least 1000 times smaller than the luminosity in {\it Punctum}. Supernova remnants (SNRs) can produce a large radio luminosity to $\sim 10^{35}$ erg/s \cite{Macias}. However, the mm polarization is usually very low for SNRs (e.g. $\sim 7-8$\% in the Crab Nebula \cite{Ritacco}). Besides, although the jets from microquasars or stellar-mass black holes can reach a luminosity of $\sim 10^{33}$ erg/s \cite{Marti}, the radio polarization is also very low (e.g. $<1$\% for microquasar SS 433 and black hole V404) \cite{Marti,Hughes}. Therefore, all of the potential models cannot provide satisfactory explanation for the radio emission features of {\it Punctum} \cite{Shablovinskaia}. 

On the other hand, some cosmological theories suggest that primordial black holes (PBHs) might have formed in the early universe due to the collapse of dense pockets of matter \cite{Carr}. Although the existence of PBHs is yet to confirm, there is good evidence supporting the existence of PBHs, including the observed merging events using gravitational waves \cite{Bird,Sasaki,Stasenko}. Furthermore, recent data from James Webb Space Telescope (JWST) have found that some supermassive black holes (SMBHs) have been formed at unexpected early epochs \cite{Napolitano,Grant,Maiolino}. The existence of PBHs can provide seeds for the unexpected fast growing of SMBHs \cite{Maiolino,Dayal}. If PBHs were formed long before matter-radiation equality as theories predict \cite{Carr}, each PBH would then accrete dark matter around itself to form a dark matter density spike with the size $<1$ pc \cite{Eroshenko,Adamek}. Previous studies have rigorously shown that the profile of the dark matter density spike around a PBH follows $\rho_{\rm DM} \propto r^{-9/4}$ \cite{Eroshenko,Adamek,Eroshenko2}. Such a high dark matter density at small $r$ would enhance the annihilation rate and emit a huge amount of high-energy particles like electrons and positrons. The produced electrons and positrons interacting with magnetic field would emit synchrotron radiation in radio bands. Since the magnetic field of an accreting PBH can be highly ordered \cite{Kenzhebayeva}, this scenario can account for the significant highly-polarized mm emission of {\it Punctum}. Besides, the self-absorption of synchrotron radiation at low frequencies and the feature of the synchrotron flux originating from dark matter annihilation can provide a self-consistent picture to account for the observed characteristics of {\it Punctum}. Therefore, in this article, we show that {\it Punctum} is likely a PBH containing a dark matter density spike. By formulating an analytic framework, we can also constrain relevant dark matter annihilation parameters using the data of {\it Punctum}.

\section{The PBH dark matter density spike model}
If PBHs were formed in the very early universe, their gravitational influence would drive the surrounding dark matter particles to decouple from the background expansion and move inwards. Finally, a dark matter density spike would be formed around a PBH. Following the standard cosmological framework, the density profile of the dark matter density spike is \cite{Adamek}
\begin{equation}
\rho_{\rm DM}(r)=0.84 \rho_{\rm eq}(GM_{\rm PBH}t_{\rm eq}^2)^{3/4}r^{-9/4},
\end{equation}
where $\rho_{\rm eq} \approx 2.1\times 10^{-19}$ g/cm$^3$ is the density of the universe at matter-radiation equality, $M_{\rm PBH}$ is the mass of the PBH, and $t_{\rm eq}=2.4\times 10^{12}$ s is the cosmological time at matter-radiation equality (assuming the Hubble parameter $h=0.7$). At small $r$, the dark matter density would be extremely high so that the self-annihilation rate would be significantly triggered. Therefore, a significant amount of dark matter would be annihilated after the cosmological time $t_0 \sim 4\times 10^{17}$ s and the central density would approach a plateau density \cite{Adamek}:
\begin{eqnarray}
\rho_{\rm max}&=&1.5\times 10^{-16}~{\rm g/cm^3} \left(\frac{m_{\rm DM}}{1~\rm GeV} \right) \left(\frac{3\times 10^{-26}~{\rm cm^3/s}}{\langle \sigma v \rangle} \right) \nonumber\\
&& \times \left(\frac{4\times 10^{17}~{\rm s}}{t_0} \right),
\end{eqnarray}
where $\langle \sigma v \rangle$ is the velocity-weighted annihilation cross section, $m_{\rm DM}$ is the dark matter mass and $t_0$ is the age of the universe. The core radius of the plateau density is
\begin{eqnarray}
r_c&\approx &0.93 \left(\frac{\rho_{\rm eq}}{\rho_{\rm max}} \right)^{4/9}(GM_{\rm PBH}t_{\rm eq}^2)^{1/3} \nonumber\\
&=&4.6\times 10^{15}~{\rm cm} \left( \frac{m_{\rm DM}}{1~{\rm GeV}} \right)^{-4/9} \left(\frac{M_{\rm PBH}}{M_{\odot}} \right)^{1/3} \left(\frac{\langle \sigma v \rangle}{3\times 10^{-26}~{\rm cm^3/s}} \right)^{4/9}.
\end{eqnarray}
For a very good approximation, we assume $\rho_{\rm DM}=\rho_{\rm max}$ when $r \le r_c$ and $\rho_{\rm DM}$ follows Eq.~(1) when $r>r_c$. 

\section{Radio emission of dark matter annihilation}
Dark matter annihilation would produce a large amount of electrons and positrons. These electrons and positrons can interact with magnetic field to produce synchrotron radiation in radio bands. Around an accreting black hole, the magnetic field strength can be as large as $B \sim 10$ G \cite{Dallilar}. Therefore, the electrons and positrons produced from dark matter annihilation would cool down quickly through synchrotron radiation without any significant diffusion. The radio flux at frequency $\nu$ can be best approximated by the following expression \cite{Bertone,Profumo}:
\begin{eqnarray}
S_{\rm DM}&=&\frac{1}{4\pi \nu D^2} \left(\frac{9\sqrt{3} \langle \sigma v \rangle}{4m_{\rm DM}^2} \right) \nonumber\\
&& \times \int dr[4\pi r^2 \rho_{\rm DM}^2E(\nu,B)Y(m_{\rm DM})],
\end{eqnarray}
where $D$ is the distance to the PBH, $E(\nu,B)=14.6(\nu/{\rm GHz})^{1/2}(B/\rm \mu G)^{-1/2}$ GeV and 
\begin{equation}
Y(m_{\rm DM})=\int_{E(\nu,B)}^{m_{\rm DM}} \frac{dN_e}{dE'}dE'
\end{equation}
with $dN_e/dE'$ being the injection spectrum of the electrons and positrons. If we consider that dark matter annihilation would primarily produce electron and positron pairs only, we have $dN_e/dE' \approx 2\delta(E'-m_{\rm DM})$ and $Y({m_{\rm DM}})\approx 2$. If $m_{\rm DM}$ is greater than the mass of a muon $m_{\mu} \approx 105.66$ MeV, then dark matter could annihilate primarily into muon pairs first and then forming a cascade of electrons and positrons. In this case, $dN_e/dE'$ would be a continuous spectrum in $E$ depending on $m_{\rm DM}$ and the value of $Y(m_{\rm DM})$ would be larger than 2. In the followings, we simply assume that dark matter annihilate primarily into electrons and positrons for illustration (i.e. $Y(m_{\rm DM}) \approx 2$). In general, our model could be extended to other possible annihilation channels and the value of $Y(m_{\rm DM})$ would be revised accordingly.

If {\it Punctum} is a PBH, taking $D=3.72$ Mpc \cite{Shablovinskaia,Koss} and $\nu=100$ GHz, we get
\begin{equation}
S_{\rm DM}=0.28~{\rm mJy} \langle \sigma v \rangle_{26}^{1/3}m_{\rm GeV}^{-4/3}M_{\rm PBH,{\odot}}B_{\rm G}^{-1/2},
\end{equation}
where $\langle \sigma v \rangle_{26}=\langle \sigma v \rangle/3\times 10^{-26}~{\rm cm^3/s}$, $m_{\rm GeV}=m_{\rm DM}/1~{\rm GeV}$, $M_{\rm PBH,{\odot}}=M_{\rm PBH}/M_{\odot}$ and $B_{\rm G}=B/1~{\rm G}$. Near an accreting PBH, the number density is extremely high so that synchrotron self-absorption would be very significant at low frequencies (e.g. $\nu < 100$ GHz) \cite{Longair}. Therefore, the observed radio flux might be smaller than the actual radio flux due to self-absorption. In active galactic nuclei (AGN), due to self-absorption, the observed flux at small frequencies would be suppressed by a positive spectral index $\alpha'$ while the observed flux at large frequencies would be close to the actual flux emission. The entire radio spectrum can be best described by the following functional form \cite{Turler,Cho}:
\begin{equation}
S(\nu)=S_m \left(\frac{\nu}{\nu_m} \right)^{\alpha'} \frac{1-\exp\left[-\tau_m(\nu/\nu_m)^{\alpha-\alpha'} \right]}{1-\exp(-\tau_m)},
\end{equation}
where $S_m$ is the observed maximum flux when $\nu=\nu_m$, $\tau_m=1.5(\sqrt{1-8\alpha/3\alpha'}-1)$ and $\alpha$ is the spectral index representing the actual flux emission. For homogeneous self-absorption, we have $\alpha' \approx 5/2$ \cite{Longair,Turler}. From Eq.~(4), we have $\alpha=-1/2$. Therefore, we get $\tau_m \approx 0.357$ and the actual flux emission at $\nu=\nu_m$ is about $1.19S_m$ (see Fig.~1). Note that the spectral shape in Fig.~1 (with self-absorption) originates from Eq.~(7). The peak value $S_m$ would be derived from the observed radio data only, but not from the properties of dark matter or density spike. The constrained $S_m$ from the observed spectrum can be used to determine the properties of dark matter and the PBH (see below).

Observations of {\it Punctum} indicate a flux of $0.104 \pm 0.018$ (S1) or $0.125\pm 0.016$ (S2) at $\nu \approx 90-104$ GHz \cite{Shablovinskaia} and there is no detection at $\nu<23$ GHz and $\nu>300$ GHz. One would expect that the observed flux at $\approx 100$ GHz might be the maximum flux $S_m$ due to self-absorption effect. The data observed are also consistent with a spectral index of zero so that the position of $\nu \approx 100$ GHz in the spectrum is close to the maximum turning point of the spectrum \cite{Shablovinskaia}. For instance, the maximum flux position for the spectrum of our SMBH Sgr A* is at $\nu_m \approx 100-300$ GHz \cite{Witzel}. If this is the case for {\it Punctum}, we can write $S_{\rm DM} \approx 1.19S_m$ at $\nu=100$ GHz and get our first constraint using Eq.~(6):
\begin{equation}
\langle \sigma v \rangle_{26}^{1/3}m_{\rm GeV}^{-4/3}M_{\rm PBH,{\odot}}B_{\rm G}^{-1/2}=0.44.
\end{equation}
Here, we have assumed $S_m=0.104$ mJy at $\nu=100$ GHz.

Furthermore, the size of the self-absorption region is given by \cite{Laor}
\begin{equation}
R_{\rm SA}=1.7\times 10^{18}~{\rm cm}~L_{30}^{1/2}B_{\rm G}^{1/4}\nu_{\rm GHz}^{-5/4},
\end{equation}
where $L_{30}=L_{\nu}/10^{30}~{\rm erg/s/Hz}$ with $L_{\nu}$ the spectral luminosity and $\nu_{\rm GHz}$ the frequency in GHz. If the magnetic field is large enough ($B=B_{\rm eq}\approx 1370$ G) to maintain the equipartition relation (i.e. the magnetic energy density = photon energy density), the size of the self-absorption would be \cite{Laor}
\begin{equation}
R_{\rm SA}=1.5\times 10^{18}~{\rm cm}~L_{30}^{0.4}L_{46}^{0.1}\nu_{\rm GHz}^{-1},
\end{equation}
where $L_{46}=L_{\rm bol}/10^{46}~{\rm erg/s}$ represents the bolometric luminosity. Since $R_{\rm SA}$ decreases when $\nu$ increases, the synchrotron flux observed would be closer to the emission flux for larger $\nu$ due to a smaller size of self-absorption region. This implies that the maximum turning point of the observed synchrotron spectrum corresponds to the dominating emission region almost equal to the size of the self-absorption region. Since $S_{\rm DM}\propto r^{-3/2}$ outside the core region, the dominating emission region of the dark matter density spike originates from the core region $r \le r_c$. Hence, we have $R_{\rm SA} \approx r_c$ at $\nu \approx 100$ GHz and this gives the second constraint for $B<B_{\rm eq}$:
\begin{equation}
M_{\rm PBH,{\odot}}^{1/3} \langle \sigma v \rangle_{26}^{4/9}m_{\rm GeV}^{-4/9}B_{\rm G}^{-1/4}=0.0016.
\end{equation}
For $B=B_{\rm eq}$, assuming $L_{\rm bol}=2\times 10^{43}$ erg/s in the central region of NGC 4945 \cite{Perez}, the second constraint is
\begin{equation}
M_{\rm PBH,{\odot}}^{1/3} \langle \sigma v \rangle_{26}^{4/9}m_{\rm GeV}^{-4/9}=0.0090.
\end{equation}

\begin{figure}
\vskip 3mm
\includegraphics[width=140mm]{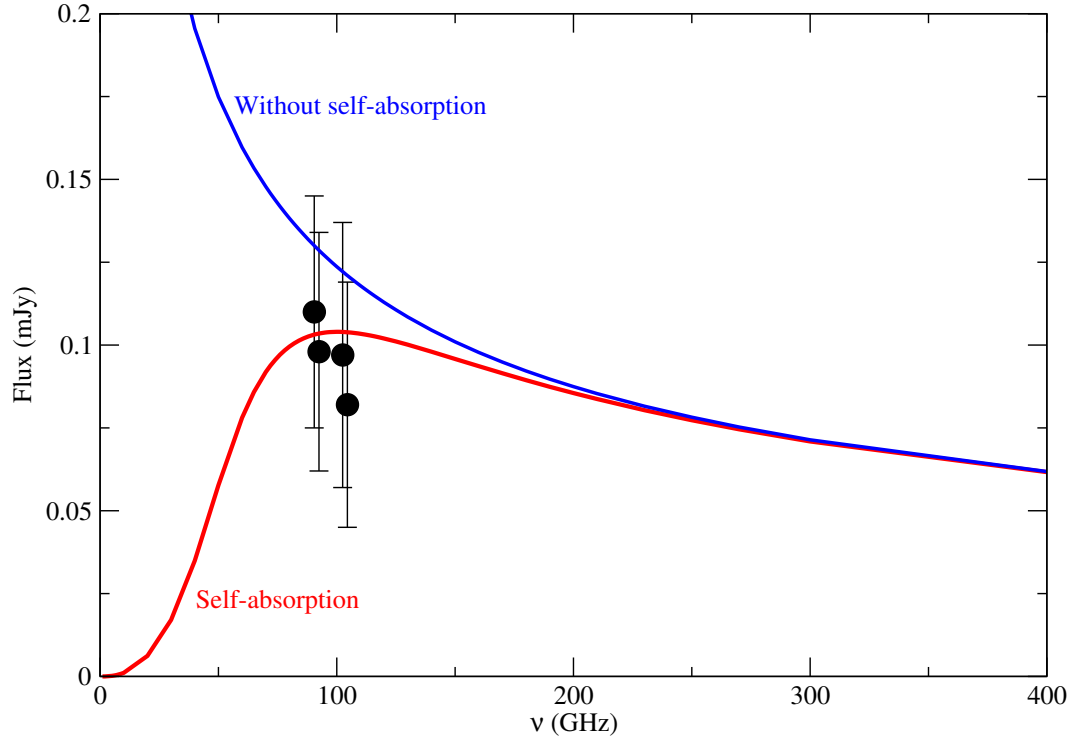}
\caption{The red line indicates the spectrum of {\it Punctum} using Eq.~(7) involving self-absorption. The blue line represents the projected emission spectrum of dark matter annihilation without self-absorption. The data are extracted from the S1 measurement of \cite{Shablovinskaia}.}
\label{Fig1}
\vskip 3mm
\end{figure}

\section{Results}
In Fig.~2, we plot the two constraints using Eq.~(8) and Eq.~(11) for a typical magnetic field $B=100$ G and two PBH masses $M_{\rm PBH}=10M_{\odot}$ and $40M_{\odot}$. The green region represents the excluded parameter space in which the peak synchrotron energy $E(\nu,B)$ is larger than the dark matter mass. For $\nu \approx 100$ GHz, the peak synchrotron energy would be $14.6$ MeV for $B=100$ G. Therefore, any dark matter mass $<14.6$ MeV would be excluded in our model because $E(\nu,B)$ must be less than $m_{\rm DM}$. The intersections between two solid lines and two dashed lines are the solutions of $\langle \sigma v \rangle$ and $m_{\rm DM}$ for the two different scenarios. For instance, if $M_{\rm PBH}=10M_{\odot}$, $m_{\rm DM} \sim 25$ MeV and $\langle \sigma v \rangle \sim 10^{-33}$ cm$^3$/s can provide a viable solution to explain the observed features of {\it Punctum}. 

In fact, on solving the equations of the two constraints, one can find that the annihilation cross section does not depend on the dark matter mass:
\begin{equation}
\langle \sigma v \rangle_{26}=\left\{
\begin{array}{ll}
9.9 \times 10^{-9}B_{\rm G}^{1/4} & {\rm for }\,\,\, B\le B_{\rm eq} \\
1.6\times 10^{-6}B_{\rm G}^{-1/2} & {\rm for}\,\,\, B=B_{\rm eq} \\
\end{array}
\right.
\end{equation}
As the dependence on $B_{\rm G}$ is insensitive, for $B\sim 1-1000$ G, we get $\langle \sigma v \rangle \sim 10^{-33}$ cm$^3$/s. Interestingly, this is close to the predicted annihilation cross section for $\sim 10$ MeV dark matter accounting for the 511 keV excess at the Galactic Center for the Gondolo-Silk dark matter density spike model \cite{Torre}. Moreover, the constrained annihilation cross section can safely satisfy all of the current bounds for MeV and sub-GeV dark matter \cite{Depta,Siegert}.

By putting Eq.~(13) into the first constraint, we can obtain another relation connecting $M_{\rm PBH}$ and $m_{\rm DM}$ for different values of $B$. In Fig.~3, we plot the relations between $m_{\rm DM}$ against $M_{\rm PBH}$ for $B$ equal to 1 G, 10 G, 100 G and the equipartition scenario. The large dot in each plotted relation indicates the lower bound of $m_{\rm DM}$ and $M_{\rm PBH}$. For example, for $B=1$ G, we have $m_{\rm DM} \ge 146$ MeV and $M_{\rm PBH} \ge 16M_{\odot}$. For PBH mass $M_{\rm PBH} \sim 10-100M_{\odot}$, the constrained range of dark matter mass is $\sim 10-1000$ MeV. 

\begin{figure}
\vskip 3mm
\includegraphics[width=150mm]{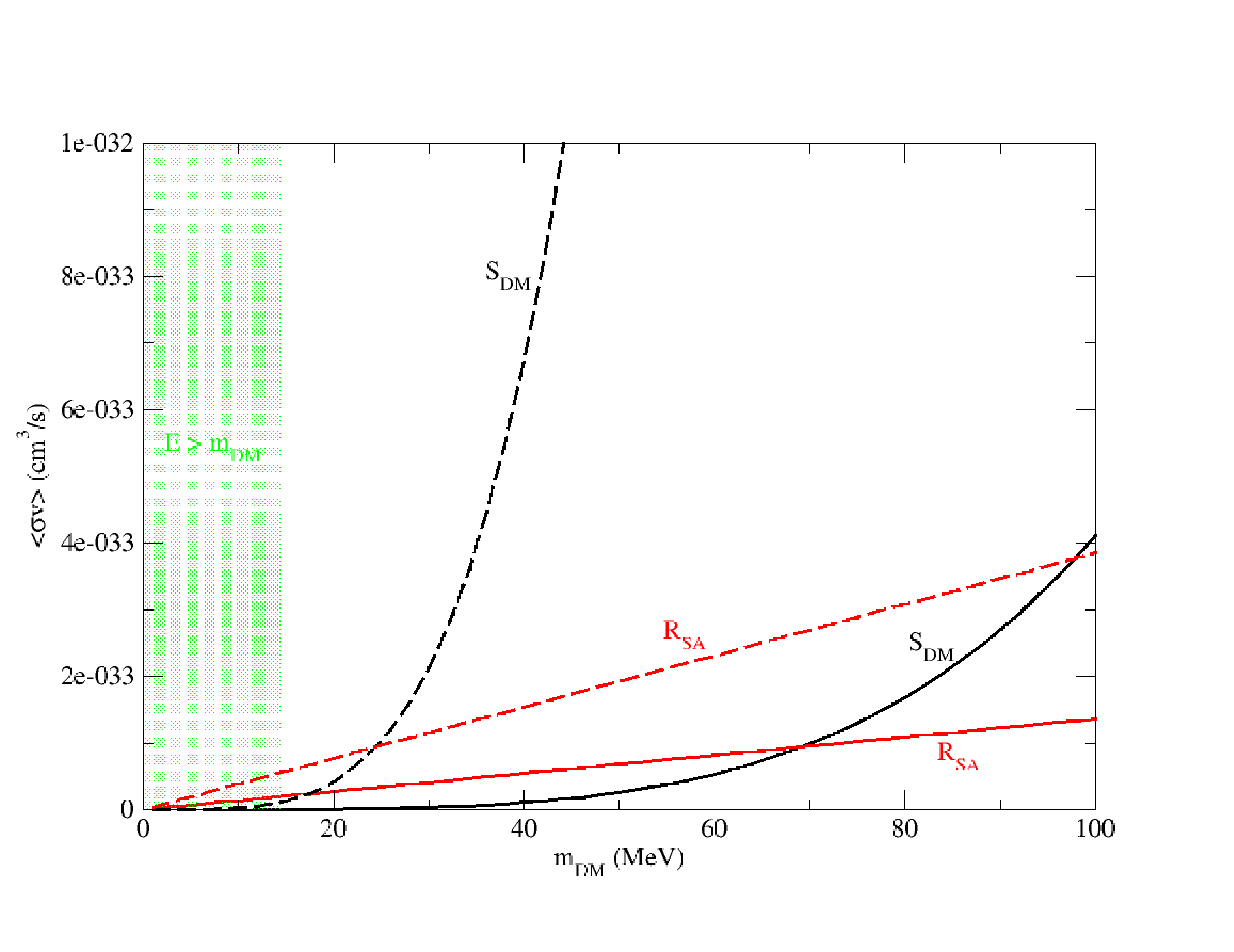}
\caption{The black lines and red lines represent the first constraint and second constraint respectively, considering the emission due to dark matter annihilation and self-absorption effect. The solid and dashed lines respectively indicate the parameters ($B=100$ G, $M_{\rm PBH}=40M_{\odot}$) and ($B=100$ G, $M_{\rm PBH}=10M_{\odot}$) assumed in the constraints. The green region is the excluded parameter for $E>m_{\rm DM}$. The intersections between the solid lines and dashed lines are the solutions of $m_{\rm DM}$ and $\langle \sigma v \rangle$.}
\label{Fig2}
\vskip 3mm
\end{figure}

\begin{figure}
\vskip 3mm
\includegraphics[width=140mm]{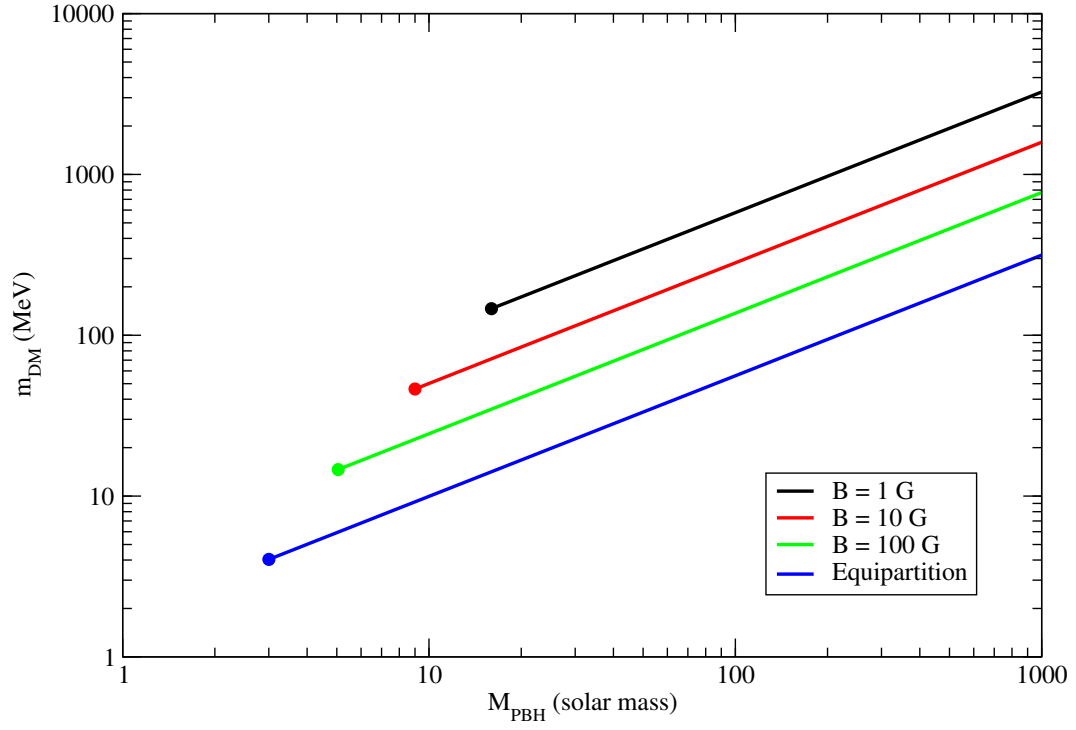}
\caption{The black, red, green and blue lines represent the relation between $m_{\rm DM}$ and $M_{\rm PBH}$ for 1 G, 10 G, 100 G and equipartition ($\approx 1370$ G) respectively. The colored large dots represent the lower limits of $m_{\rm DM}$ and $M_{\rm PBH}$ for the corresponding magnetic field strength.}
\label{Fig3}
\vskip 3mm
\end{figure}

\section{Discussion}
Recent discovery of {\it Punctum} from ALMA has shown some mysterious features, in which no known categories of objects can offer satisfactory explanation \cite{Shablovinskaia}. In this article, we discuss an innovative proposal that {\it Punctum} is indeed a PBH with a dark matter density spike. The synchrotron radiation originating from dark matter annihilation in the core region of the dark matter density spike can provide large enough energy to account for the observed radio signal at 100 GHz. Since the magnetic field can be highly ordered around an accreting black hole \cite{Kenzhebayeva}, it can explain the highly polarized signal $\sim 50$\% (some unknown thermal noise or imperfect alignment of magnetic field might exist). Besides, the emission mainly originates from the core region of the dark matter density spike $r \le r_c$. For a possible set of parameters $M_{\rm PBH}=10M_{\odot}$, $m_{\rm DM}=25$ MeV and $\langle \sigma v \rangle \sim 10^{-33}$ cm$^3$/s, using Eq.~(3), we have $r_c \sim 10^{-5}$ pc. Therefore, the compactness of the dark matter density spike is also consistent with the upper limit of the size of {\it Punctum} ($\le 2$ pc or 0.11 arcsec) \cite{Shablovinskaia}. Therefore, this proposal can match all of the observed features of {\it Punctum} and does not have any difficulties.

As the observational data are consistent with the synchrotron self-absorption scenario, we can determine the constraints of dark matter mass $m_{\rm DM} \sim 10-1000$ MeV and annihilation cross section $\langle \sigma v \rangle \sim 10^{-33}$ cm$^3$/s based on the benchmark ranges of magnetic field $B \sim 1-1000$ G and PBH mass $M_{\rm PBH} \sim 10-100M_{\odot}$. In particular, for $m_{\rm DM} \sim 10$ MeV, the order of magnitude of the constrained annihilation cross section is somewhat close to the one invoked to explain the Galactic Center 511 keV excess signals, assuming the Gondolo-Silk dark matter density spike model \cite{Torre}. Therefore, our results provide an additional support to change our attention of investigating weakly interacting mass particle (WIMP) dark matter mass from GeV to MeV or sub-GeV scale. The small annihilation cross section may also suggest a non-standard origin of dark matter production in our universe as the standard paradigm suggests dark matter formed thermally with $\langle \sigma v \rangle \approx 3 \times 10^{-26}$ cm$^3$/s \cite{Steigman}.

It has been suggested that the existence of PBHs can provide seeds for SMBH growing at early epochs \cite{Maiolino}. Recent observations from gravitational waves generally favor the existence of PBHs \cite{Bird,Sasaki,Stasenko}. Some studies even suggest that the nearby two black hole systems, A0620-00 and XTE J1118+480, are PBHs because of the possible existence of dark matter density spikes \cite{Chan,Ireland}. The alleged dynamical origin of the Gaia BH3 binary also suggests that the black hole Gaia BH3 may be a PBH \cite{Bhalla}. Traditionally, we identify a black hole being a PBH candidate if its mass falls within the forbidden mass gap of black holes (i.e. $\sim 2-5M_{\odot}$ or $>50M_{\odot}$) \cite{Bi}. However, it is possible that the mass of a PBH lies outside the forbidden mass gap of black holes, despite the actual mass gap is still uncertain \cite{Croon}. In this study, based on the fact that a normal stellar-mass black hole is hard to form a dark matter density spike \cite{Ireland}, we have demonstrated a completely new way to identify PBHs by determining whether they contain dark matter density spikes. If there are PBHs containing dark matter density spikes, dark matter annihilation would give electrons and positrons to generate synchrotron radiation peaked at $\sim 100$ GHz. Therefore, we expect that future millimeter observations might be able to identify more objects like {\it Punctum}. This would initiate a new way to search for PBHs and constrain dark matter properties at the same time. 

\section{Acknowledgements}
The work described in this paper was partially supported by the grant from the Research Grants Council of the Hong Kong Special Administrative Region, China (Project No. EdUHK 18300324).

\section{Data availability statement}
The data underlying this article will be shared on reasonable request to the corresponding author.





\end{document}